\documentclass[prb,preprint,showpacs]{revtex4}
\usepackage[isolatin]{inputenc} 
\usepackage[T1]{fontenc} 
\usepackage{times}
\usepackage[pdftex]{graphicx} 

\begin{document}
\unitlength1cm
\author{M.~Letz$^{(a)}$, L.~Parthier$^{(b)}$}

\affiliation{$^{a}$ Schott AG, Research and Development, D-55014~Mainz, Germany}
\affiliation{$^{b}$ Schott Lithotec AG, Otto-Schott Str. 13,
D-07745~Jena, Germany}

\title{Charge centers in CaF$_2$: Ab initio calculation of elementary
physical properties.}

\begin{abstract}
Charge centers in ionic crystals provide a channel for elementary
interaction between electromagnetic radiation and the lattice. We
calculate the electronic ground state energies which are needed to
create a charge center -- namely a $F$- and a $H$-center. In well agreement
with common understanding the $F$-center results in being accompanied by a
small lattice distortion whereas the $H$-center is accompanied by a very
large lattice deformation. Opposite to the common understanding the
additional positive charge in the charge center results rather to be
localized on a F$_4^{3-}$ complex than on a F$_2^-$-complex.
From the ground states of the charge centers we
derive binding energies, diffusion barriers and agglomeration energies
for $M$-center formation. 
These microscopic quantities are of fundamental
interest to understand the dynamic processes which are initiated if
the crystals interact with extreme intense deep ultra violet
radiation. We further derive the equilibrium concentrations of charge
centers in grown crystals. 
\\[0.5cm]
\end{abstract}

\date{\today}
\maketitle   


\section{Introduction}
\label{sec:intro}

Calciumfluoride (CaF$_2$) is a key material for building refractive
optical elements in the deep ultra violet (DUV) range of the
electromagnetic spectrum. Especially for structuring microchips an
optical system in the DUV range is of extreme technical importance. 
The reason for this is that the smallest possible structure size, which
can be obtained by a microlithographic process is proportional to the
wavelength of the radiation used for structuring. 
In
the last years immense effort has been made in order to provide large
single crystals of CaF$_2$ with high optical quality. These
microlithographic processes are performed at wavelengths of excimer
laser radiation sources. These are mainly 193nm (ArF excimer laser),
157nm (F$_2$ excimer laser) and 248nm (KrF excimer laser). CaF$_2$ is
chosen as an optical material due to its cubic crystal structure with
(nearly \cite{burnett2001, letz2003b}) perfect optical isotropy,
due to its chemical durability and its mechanical properties which
make it applicable for lens fabrication.     

Recently high purity single crystals have been grown with
impurity concentrations below the ppm level. However, even in these
pure single crystals a finite probability for absorption processes
remains. Since any absorption process allows for a deposition of
energy in the crystal, any energy deposition will in turn lead to a
change of material properties. Therefore there exists strong interest
to further our understanding of the microscopic mechanism beyond UV absorption
in fluoride crystals. 

In the past there have been a large number of investigations on charge
centers in fluoride crystals in the early 70th (see e.g. \cite{hayes1974,
 fowler1968}). One of the main driving forces at that time was to find
materials for color center lasers and hosts which show zero phonon
lines for spectral hole burning. Further on CaF$_2$ is used as a model
system for ionic conductivity since the F-sub lattice melts
approximately 100K below the melting point of the crystal. Since that
time the quality of the crystals was increased dramatically to to the
needs of the semiconductor industry for refractive optical materials
in the deep UV wavelength range.  

Even in an ideal material, where no point defects are present,
electromagnetic radiation can be absorbed with a small but finite
probability via a two photon process \cite{tsujibayashi2000,goerling2005}. Such an
absorption process will 
therefore quadratically increase with increasing radiation
intensity and becomes of strong importance with increasing
radiation intensity. Such a two photon process will separate charges and will
create an electron--hole pair, provided the sum of the energy of the
two photons exceeds the 11.2 eV of the excitonic two particle bound
state of the $\Gamma$--exciton \cite{tomiki69}. Above this energy a large probability
for absorption is present \cite{barth90}. The charges of the electron and the
hole have further the possibility to lower their energy by distorting
the surrounding lattice. This means that a localization of the
electron--hole pair takes place together with a lattice
distortion. The resulting state is called an $F$-$H$-pair
\cite{williams1976,lindner01} or a self trapped exciton (STE) \cite{mizuguchi1999}. This
$F$-$H$-pair can recombine with an optically forbidden transition, which
leads to the fluorescence at 278nm with a relatively large lifetime at
room temperature of 1.1$\mu$s \cite{williams1976,tanimura2001,muehlig2002,cramer2005}.   

In a real material there are a number of defects present. In this case
the defect concentration is governed by entropy. The concentration at
room temperature is in this case not necessarily determined by its
room-temperature equilibrium value. This is due to the fact that the
system falls with respect to defect statistics out of equilibrium
already at the crystallization point where defect diffusion becomes
strongly hindered.

At high radiation rates the creation of $F$-$H$-pairs provides the major
source towards radiation dependent change of optical material
properties of CaF$_2$. Due to thermal diffusion the charge centers can
move around in the material. Above 170 K \cite{170K} the charge
centers are known to be mobile. This mobility increases on one site the
probability of recombination of $F$-- and $H$--centers. On the other site
mobile charge centers can be trapped by point defects and impurities
in the material which can lead to room temperature stable charge
centers. These temperature dependence is a strong hint that activation
energies for diffusive transport of charge centers are of crucial
importance when understanding radiation dependent change in material
properties of CaF$_2$. In the present paper we calculate energies for
charge center formation and derive the necessary microscopic
parameters for charge center diffusion.

The paper is organized as follows: In section \ref{sec:interact} we
phenomenologically describe the interaction between radiation and the
fluoride crystal. In section \ref{sec:band} we discuss the band
structure of CaF$_2$ which is needed as a starting point when
investigating the localized charge centers. In the next section
\ref{sec:struct_f_h} we report calculation on a larger unit cell up to
Ca$_{108}$F$_{216}$ where either one fluor ion is removed and an electron
is localized on the vacancy or where an additional fluor ion has been
added forming an essentially positively charged complex where the hole
is localized. The lattice deformation is discussed in subsection
\ref{sec:latt_relax} and the obtained electronic structure in
\ref{subs_f} and \ref{sec:h}. In section \ref{sec:equ_f_h} we calculate
the equilibrium concentrations of $F$- and $H$-centers in an as grown crystal.

\section{Interaction of UV radiation with CaF$_2$}
\label{sec:interact}

For a qualitative understanding of the dynamic processes involving
charge center formation and charge center motion we follow the
arguments of \cite{kuzokov1998}. 
We assume that a two photon
process creates a pair of an $F$-center and a $H$-center which are
spatially uncorrelated. There is further no long range interaction
between these pairs an assumption which is physically justified by the
screening of the coulomb potential.

Further all interaction between the $F$ and $H$ centers is neglected. That
means we assume an uniform density for both centers. This
approximation gets exact in the dilute limit. This means that the two
particle correlation functions $g_{\kappa,\kappa'}(r)$ is assumed to
have the uniform value $1$. 
As the only driving
force for $F$-$H$-pair recombination we assume statistical fluctuations of
the charge center positions.
\begin{equation}
\rho^{(2)}_{\kappa,\kappa}(r) = \rho^{(1)}_{\kappa} \;
g_{\kappa,\kappa}(r) \approx \rho^{(1)}_{\kappa} \;\;\;\;\; \kappa
\in \{F,H\}
\end{equation}
and for the non-diagonal correlation function we assume:
\begin{equation}
\rho^{(2)}_{F,H}(r)=\rho^{(2)}_{H,F}(r) \approx \sqrt{ \rho^{(1)}_{F}  \rho^{(1)}_{H}}
\end{equation}
With the above approximations we can directly write down a system of
rate equations for the creation of $F$ and $H$ center in the ideal
material. 
\begin{equation}
\label{eq:id_sys}
\frac{\partial \rho^{(1)}_{\kappa}}{\partial t} = I^2 p - K \;  
\sqrt{ \rho^{(1)}_{F}  \rho^{(1)}_{H}}
\;\;\;\;\; \kappa \in \{F,H\}
\end{equation}
The first term on the right site creates pairs of $F$- and $H$ charge
centers. This creation goes via a two photon process and is therefore
proportional to the squared radiation intensity $I^2$. The second term
on the right site of eq. (\ref{eq:id_sys}) describes the recombination
of particles. This recombination is proportional to the probability of
finding two particles within a distance smaller than  the
recombination radius $r_0$ and is proportional to the effective
recombination rate $K$.
\begin{equation}
K=4 \pi r_0 (D_F +D_H)
\end{equation}
Here $D_F$ and $D_H$ are the diffusion constants for $F$-center and
$H$-center diffusion. The temperature dependence of these diffusion
constants is known from \cite{atobe1979} as:
\begin{equation}
D_{\kappa} = d_{\kappa} e^{-\frac{E_{\kappa}}{k_B T}}
\end{equation}
with $E_F=0.7$ eV and $E_H=0.46$ eV.
In order to obtain microscopic information on the material parameters
needed for charge center diffusion, we do an ab initio calculation of
the charge centers together with the resulting lattice distortions.
Among other results we also confirm a well known result \cite{hayes1974}
that there is a strong asymmetry between the number of $F$-centers and
$H$-centers in CaF$_2$. At the melting point of the F-sublattice, where
the system falls out of equilibrium with respect to its defect
concentration there is already a reasonable number of $F$-centers
present but there are practically no $H$-centers (see sec. \ref{sec:equ_f_h}).

\section{Method of calculation}
\label{sec:meth}

In the present work we use a commercially available density functional package
(vasp \cite{vasp}). The code is a plane-wave code which uses ultrasoft
preudopotentials. The generalized gradient approximation (GGA) is used to
calculate the exchange correlation energy. 
Through most of the calculation we set the plane wave cutoff for the
numerical evaluation of the electron density to 500 eV. This was due
to restriction of computer time. In appendix \ref{sec:convergence} we
show that this choice is in general sufficient. For the larger clusters with
32 Ca ions and 64 F ions in
section \ref{sec:struct_f_h} we perform a $\Gamma$--point calculation, which
corresponds to a k-mesh of 0.575 \AA $^{-1}$. The criterion for the
electronic convergence was set to $\Delta E / E = 10^{-6}$.

\section{Band structure of CaF$_2$}
\label{sec:band}
As a starting point the band structure of CaF$_2$ has been calculated.
The resulting bandstructure is in well agreement with
experiments and other ab initio calculations \cite{kim2004}. As expected for a
pure DFT calculation the resulting band gap is with 7.7eV too small
compared with the 11.8eV which are measured experimentally.  

A screened exchange calculation improves the too small band gap. 
\cite{kim2004} Also a similar calculation which uses an interpolation between
DFT  and Hartree--Fock approximation \cite{shi2005} arrives at a similar
result. Details on the band structure together with a
calculation of phonon modes which are compared to neutron scattering
results can be found in \cite{schmalzl2003}. In \cite{khenata2005} also
mechanical properties of CaF$_2$ and other fluorites are calculated.

\section{Structure of $F$- and $H$-center}
\label{sec:struct_f_h}

In order to calculate charge centers we
set up a cluster of Ca$_{32}$F$_{64}$ using periodic boundary
conditions. Due to the large size of the elementary cell calculations
are performed using one k-point only ($\Gamma$-point calculation). 
Removing one F atom and allowing for a relaxation of
the structure into the energy minimum gives an $F$-center. Adding an F
atom and allowing for structural relaxation is a configuration of a
$H$-center. For the structural relaxation, which takes account of the lattice
deformation, the size of the elementary cell was kept fix.
In order to obtain charge neutrality in the clusters an
extra positive charge is trapped near the additional F ion and an
extra negative charge is trapped in the F-ion vacancy.
Due to the periodic boundary conditions a periodic array of
charge centers is calculated where the distance is equal to four
lattice constants. Since the electronic wave functions of such local
defects drop of exponentially there should be a small but finite
interaction between the charge centers. A check for finite size
effects due to defect--defect interaction is made using
larger systems of Ca$_{108}$F$_{216}$ cluster and shown in section \ref{sec:finite_size}.
 
\subsection{$F$$^-$-center}
\label{subs_f}

Removing one F-atom leads to a cluster with 
Ca$_{32}$F$_{63}$. 
Since the F-atom had an electron closely attached
to it, this electron is now localized in the F--vacancy. A defect
level with an exponentially decaying wave function is
created in the band gap, where the chemical potential moves in. This leads
to a half filled single electron level, where the electron can be
excited with a photon leading to an absorption in the optical
transmission spectrum. Such a half filled state
has a spin of $S=1/2$  which makes it possible for spin resonance
measurements \cite{bericht_spaeth} to observe the defect. As a next step a relaxation of the
structure is performed. As a result there is only minor change in the
structure. The electronic band structure of the resulting defect is
plotted in Fig. \ref{fig:fig2}. The resulting change in electronic
density compared to the undisturbed configuration is shown in
Fig. \ref{fig:fig3}. The additional charge is well localized at the
defect and only weak lattice distortion is seen. The symmetry of the
$F$-center is tetrahedron symmetry T$_d$- In appendix \ref{sec:group} we
calculate the splitting of excited energy levels due to the tetrahedron
symmetry using arguments from group theory.

\subsection{$M$--center}

A pair of two $F$--center is called an $M$-center. Usually entropy
will be the dominant driving force for a homogeneous distribution of
$F$--centers. On the other hand the overall lattice distortion which is
needed to trap an electron as a $F$--center can be reduced if two
$F$--center join forming an $M$--center. By removing a second
F$^-$--ion from our cluster with periodic boundary conditions we
create an $M$--center. herby the two $F$--center can be oriented along
different directions in the crystal. We investigate $M$--center in the
100 direction, the 110 and 111 direction. In the latter case we have
two possibilities. One where a Ca$^{2+}$ is in between the two missing
F$^-$--ions and the second one with an empty space in between. By
calculating the difference in total energy between an $M$--center
(Ca$_{32}$F$_{62}$) and a 
Ca$_{32}$F$_{64}$ cluster in comparison with twice the total energy
of a $F$--center configuration we obtain an expression
for the agglomeration energy of two $F$--center forming an $M$--center.
\begin{equation}
\Delta E_{M-aggl} = E_{Ca_{32}F_{62}}  +  E_{Ca_{32}F_{64}} - 2 E_{Ca_{32}F_{63}}
\end{equation}
We can think of this configuration as taking two $F$--center
infinite distance away from each other and comparing this state with
two $F$--center on nearest (or next-nearest or next-next-nearest)
neighbor sites.
The result is summarized in table \ref{tab:m-cent}. The configuration
where the two $F$-center are on nearest neighbor sites orientized along
the 100 direction is the 
most preferred one. 

\subsection{$H$-center}
\label{sec:h}

For a configuration which is used as a starting point for the
$H$$^+$--center, a F-ion is replaced by a pair of F-ions where
according to \cite{tanimura2001} the line connecting the two F-atoms
points along the 110 direction. 
In this way a Ca$_{32}$F$_{65}$--cluster is obtained.
After a relaxation of the structure a
stable configuration with localized defect level in the band gap is
obtained. The resulting band structure is plotted in
Fig. \ref{fig:fig4}. The charge distribution is plotted in
Fig. \ref{fig:fig5}. The additional positive charge is located in the
center of four F$^-$ ions. In this way a F$_4^{3-}$ complex is
created. This is opposite to the standard literature e.g. \cite{hayes1974,
pick1972} were the $H$--center is described as a F$_2^-$
complex. The result of the calculation is very clear therefore there
is little doubt on the quality of the result. In the 70'th when
the microscopic models of the charge centers in fluoride crystals have
been developed, there was no possibility to perform such a detailed
microscopic calculation. The details of the geometry of the four
F-ions which form the F$_4^{3-}$ center is plotted in fig. \ref{fig:fig6}. 
One can further see that the $H$-center is accompanied by a very large
lattice distortion. This means that a large part of the energy, which
is needed to create a localized but unbound electron--hole pair
originates from lattice  distortion around the $H$-center. 
Therefore the main part of the energy which is
gained by trapping an electron--hole pair as a self--trapped exciton
(STE) comes from the release of the lattice distortion around the
$H$-center when a $F$-center is close.
In addition
this is the reason, why it was not possible in the past to observe
single $H$-centers. Usually a crystal is cooled down from the melting
point and falls out of thermodynamic equilibrium with respect to its
defect concentration at a temperature which is shortly below the
melting point. At that temperature a large number of F$^-$ vacancies
is present but only a very small number of interstitial F-ions. This
means that the crystal even at the slowest cooling rates  never
reaches its thermodynamical equilibrium 
at zero temperature, where each F$^-$ position is occupied by a
F-ion. In fact there will be a reasonably large number of F$^-$
vacancies. By heating the crystal and quenching it down to liquid
nitrogen temperatures it is possible to additively color the system by
creating $F$-centers. This argument is quantified in more detail in
sec. \ref{sec:equ_f_h}. A defect which is discussed in CaF$_2$
as well is the v$_k$-center. In the v$_k$-center two F$^-$-ions are
shifted out of their equilibrium positions forming an
F$_2^-$-molecule. We did not find a stable configuration for such a
bare center and believe that it is only stabilized in the vicinity of
dopant ions.

\subsection{lattice relaxation}
\label{sec:latt_relax}
In the following the energy to create an electron--hole pair in the
form of an $F$-$H$-pair is calculated. this is the total energy of an
$F$-$H$-pair. The undisturbed lattice has a total 
energy of -17.696 eV per CaF$_2$ unit. For a Ca$_{32}$F$_{64}$ this
is $-17.64\;\; 32 = -564.63$ eV. A $F$-$H$-pair which has infinite distance
is the electronic and fully converged structure of a Ca$_{32}$F$_{63}$
cluster plus the structure of a Ca$_{32}$F$_{65}$ cluster. 
\begin{eqnarray}
\Delta E_{F-H} &=& 2 E_{Ca_{32}F_{64}} - E_{Ca_{32}F_{63}} - E_{Ca_{32}F_{65}} \nonumber \\
&=&  2 \;\; 564.63 - (-556.01) - (-565.08) eV \nonumber \\
&=& -8.17 eV 
\end{eqnarray}
via an optically forbidden transition the $F$-$H$-pair can recombine with
the relative long lifetime of 1.1 $\mu$s and
finally relax into the undisturbed configuration. The STE fluorescence
which corresponds to the $F$-$H$-pair recombination is known to be at 278
nm \cite{muehlig2002}. In Fig. \ref{fig:fig7} we show a schematic plot of
the energy scheme which relates the measured 278 nm which is a photon
energy of 4.463 eV to the calculated energy difference of 8.17 eV.

\subsection{Equilibrium concentrations of charge centers in a grown
  crystal}
\label{sec:equ_f_h}
In subsections \ref{subs_f} and \ref{sec:h} we saw that there is a
strong asymmetry in the energy which is needed to produce a $F$- or
$H$-center. The main reason for this asymmetry lies in the strong
lattice distortion which results when adding a F-ion into an
interstitial lattice site in order to create a $H$-center. We now
estimate the concentration of $F$- and $H$-center which is obtained when
the system falls out of equilibrium. Therefore we compare the total
energy of Fluorine in the four possible configurations. (i) as Flour
in the regular CaF$_2$ lattice, (ii) as Flour in an interstitial
lattice site, (iii) the energy gain which is obtained when removing a
F$^-$ ion in order to create an $F$-center and (iv) the Fluor in the
F$_2$ atom of gaseous Fluor. From section \ref{sec:latt_relax} we know
that the energy which is needed to create a $F$-$H$-pair is 8.17 eV. We
now ask how this energy is distributed between the $F$- and the
$H$-center. For the $F$-center we obtain:
\begin{eqnarray}
\label{eq:def}
\Delta E_{F} &=& \left ( E_{Ca_{32}F_{64}} - \frac{1}{2} E_{F_2} -
E_{Ca_{32}F_{63}} \right ) - \Delta E_{F-H} \nonumber \\
&=&  \left (-564.63 -(-1.79) - (-556.01) \right ) - 8.17  eV \nonumber \\
&=&  1.35 eV 
\end{eqnarray}
Which is the energy gain when creating the $F$-center and
forming half of a F$_2$ molecule. 
In \cite{shi2005} a much larger value ($\approx$7.9 eV) is obtained
for $F$$^-$--center formation by comparing to the energy of a bare F
atom instead of a F$_2$ molecule. 
This is
in agreement with older ab-initio studies \cite{puchina1998}, where
small clusters of only 18 ions where studied and a diffusion
activation energy for the $F$-center of 1.69 eV results. 
For the $H$-center we obtain in the similar way:
\begin{eqnarray}
\label{eq:deh}
\Delta E_{H} &=&  \left ( E_{Ca_{32}F_{64}} + \frac{1}{2} E_{F_2} -
E_{Ca_{32}F_{65}} \right ) - \Delta E_{F-H}\nonumber \\
&=&  \left ( -564.63 + (-1.79) - (-565.08) \right ) - 8.17 eV \nonumber \\
&=&  6.82 eV 
\end{eqnarray}
The sum of both energies gives 8.023 eV which is the 
energy needed to create an $F$-$H$-pair.
In principle one should add in Eq. (\ref{eq:def}) and subtract in
Eq. (\ref{eq:deh}) the chemical potential of a corresponding F$_2$ gas phase,
which in general depends on temperature and pressure $\Delta E_F \rightarrow
\Delta E_F + \mu(T,p,N)$ and $\Delta E_H \rightarrow
\Delta E_H - \mu(T,p,N)$. For simplicity and due to a lack of knowledge of the
detailed environment, we set this term to zero. 
The point where the system falls out of equilibrium with respect to
its F-ion concentration is given by the melting temperature of the
Flour sub-lattice in CaF$_2$. This occurs at approximately 1500K which
is 150K below the melting point of the crystal. At this temperature
the equilibrium concentration of $F$-center is:
\begin{equation}
\langle n_{eq}^F \rangle = \frac{1}{v_0} e^{-\frac{\Delta E_F}{k_B T}}
  = 2.5 \; 10^{23} \; m^{-3} = 2.5 \; 10^{17} \; cm^{-3}
\end{equation}
and for the $H$-center also at a temperature of 1500 K it is:
\begin{equation}
\langle n_{eq}^H \rangle = \frac{1}{v_0} e^{-\frac{\Delta E_H}{k_B T}}
  = 1.2 \; 10^{4} \; m^{-3} = 0.012 \; cm^{-3}
\end{equation}
where $v_0=a^3=(5.46\; 10^{-10})^3 m^3 = 1.67 \; 10^{-28} m^3$ is the
volume of the elementary cell. This result  means that in ideally grown crystal there is already a strong
asymmetry between Fluor vacancies and Fluor interstitial. There are
practically no $H$-centers but already a finite number of $F$-centers.
Using the molar weight, the density of CaF$_2$ and the Avogadro
constant, $\mbox{NA}$, we can calculate the molar concentration of charge
centers in thermal equilibrium at different temperatures and as well
an absolute  particle number in $ppm$. This is done in table \ref{tab:tabfhconc}.
To calculate the equilibrium densities in $ppm$ we used
\begin{equation}
\langle n^{H/F}_{eq.}\rangle ^{ppm} = \langle n^{H/F}_{eq.}\rangle
\frac{u_{CaF_2}}{\rho \, \mbox{NA}}
\end{equation}
where $u_{CaF_2} = 78.08$ g/mol is the molar weight of CaF$_2$, $\rho
= 3.31$ g/cm$^3$ the density of CaF$_2$ and \mbox{NA} the Avogadro number.

There is even a very small but finite change in the density of the
material due to the equilibrium concentration of F-vacancies. This
change in density is:
\begin{eqnarray}
\Delta \rho &=& 19.00\, \mbox{g/mol} \, \frac{ \langle n_{eq}^F \rangle }{\mbox{NA}}
\nonumber \\
&=& 19.00 \, \frac{2.5 \; 10^{17}}{6.022 \; 10^{23}} \, \mbox{g/cm$^3$} \nonumber \\
&=& 7.9 \; 10^{-6} \, \mbox{g/cm$^3$} = 7.9 \; 10^{-3} \, \mbox{kg/m$^3$}
\end{eqnarray} 
From that follows that an extremely inhomogeneous distribution of charge
centers can in principle have an influence on the density homogeneity
of the material and therefore also on the refractive index
homogeneity. Since, however, the bare charge centers are mobile at
room temperature, this process will lead to a homogeneous distribution
already. 

\subsection{Finite size effects}
\label{sec:finite_size}

In order to estimate finite size effects we configured a larger
cluster with Ca$_{108}$F$_{216}$ for the unperturbed
configuration. In this cluster the resulting configuration for the
$F$-center and for the $H$-center from the smaller clusters was
inserted. For these calculations we restricted ourself to a plane wave
cutoff of 400 eV only due to limitations of computer power.
After this first one electronic configuration was performed and
the forces on the ions where calculated. As a result there is still a
non negligible finite size effect present. For the $F$-center we obtain:
\begin{eqnarray}
\Delta {E^F_{fin-size}}^{elec-relax} &=& 76 E_{CaF_2} + E_{Ca_{32}F_{63}} -
E_{Ca_{108}F_{215}}^{elec-relax} 
\nonumber \\
&=& -1902.59 - (-1908.57) \nonumber \\
&=& 5.98 eV 
\end{eqnarray}
If we allow for a structural relaxation in the larger cluster we get:
\begin{eqnarray}
\Delta {E^F_{fin-size}}^{struct-relax} &=& 76 E_{CaF_2} + E_{Ca_{32}F_{63}} -
E_{Ca_{108}F_{215}}^{struct-relax} \nonumber \\
&=& -1902.59 - (-1910.42) \nonumber \\
&=& 7.83 eV 
\end{eqnarray}
Where $E_{CaF_2}$ is the total energy of a single CaF$_2$ unit cell
and $E_{Ca_{32}F_{63}}$ is the total energy of the cluster from subsection
\ref{subs_f} and $E_{Ca_{108}F_{215}}$ is the total energy of the
large superstructure. 
The differences in energy result from an interaction between a single
charge center and its neighbors due to the periodic boundary
conditions of the underlying system. The major part (5.98 eV) stems
from the electronic interaction wheres only a small part (7.83 eV-
5.98 eV = 1.85 eV) stems from an interaction between the different
lattice distortions around each individual charge center.
Therefore the
major contribution of the relatively large energy difference originates
from the long range Coulomb interaction between the charge center and
it's neighbors due to the periodic boundary conditions.    

For the $H$-center we obtain from a similar calculation for the
electronic part only:
\begin{eqnarray}
\Delta {E^H_{fin-size}}^{elec-relax} &=& 76 E_{CaF_2} + E_{Ca_{32}F_{65}} -
E_{Ca_{108}F_{217}}^{elec-relax} \nonumber \\
&=& -1911.67 - (-1912.82) \nonumber \\
&=& 1.15 eV 
\end{eqnarray}
And for the structural relaxation we get:
\begin{eqnarray}
\Delta {E^H_{fin-size}}^{struct-relax} &=& 76 E_{CaF_2} + E_{Ca_{32}F_{65}} -
E_{Ca_{108}F_{217}}^{struct-relax} \nonumber \\
&=& -1911.67 - (-1913.13) \nonumber \\
&=& 1.39 eV 
\end{eqnarray}
In this case as well the electronic part (1.15 eV) as well as the
structural part due to the interaction between lattice distortions
(1.42 eV - 1.15 eV = 0.27 eV).

The reason why finite size effects are much smaller for the $H$-center
is due to the more efficient screening of the Coulomb charges in the
case of the $H$-center. In this case there is a much larger but also
much stronger localized lattice
distortion which leads to more strongly localized electronic
wavefunctions of the defect level. This can be also seen by the weak
but still present dispersion of the defect level in Figs. \ref{fig:fig2}
and \ref{fig:fig4}.



\section{conclusion}
\label{sec:}

In the present paper we calculate the ground state of the
trapped electron (bare $F$-center) and of the trapped
hole (bare $H$-center) in CaF$_2$. These charge centers are the elementary
excitations which are needed to understand the interaction between
CaF$_2$ and DUV radiation. Since DUV radiation can even in an ideal
crystal excite electron-hole pairs via a two photon process, the
electron or the hole can localize, gaining energy via a lattice
distortion. The formation energy for a $F$-$H$-pair is, however, unevenly
shared between the electron and the hole. This is the reason why
already in an as--grown--crystal a finite number of $F$-centers of the
order of 5 ppm is present, while there are no $H$-centers present. This
explains why in Fluorite crystals there are $F$-centers observed but
$H$-centers are observed only if they are stabilized by
impurities. Calculating the lattice distortion around the $F$-center, we
found no surprise. However when investigating the lattice distortion
around the $H$-center we found that the state of smallest total energy
is a configuration where the direction connecting the two interstitial
F ions points along the 110 direction of the crystal. Further the
charge is distributed rather around in a F$_4^{3-}$ complex rather than in a
F$_2^-$ complex. For the important problem of radiation stability of
CaF$_2$ under intense excimer laser radiation, the trapping of these
charge centers on impurity ions has to be considered. The radiation
stability of CaF$_2$ is of extreme importance for the optical
microlithography for structuring semiconductors.

\begin{appendix}

\section{Test of numerical accuracy due to plane wave cutoff}
\label{sec:convergence}
In a strongly ionic crystal like CaF$_2$ there is a large spatial
fluctuation of the electron charge. Therefore it is expected that a
large number of plane waves (in k-space) has to be used to achieve a
numerical accuracy which allows quantitative predictions.

In Fig. \ref{fig:conv} we have plotted the total energy of a
CaF$_2$ system as a function of the plane wave cutoff. In
Fig. \ref{fig:latt_cutoff} we show the resulting lattice constants
after a structural relaxation.  For cutoff energies of 500 eV and
larger the system starts to saturate. To minimize calculation time we
have used a plane wave cutoff of 500 eV throughout the calculations of
the larger clusters for the calculation of charge centers and
impurities in CaF$_2$. In Fig. \ref{fig:latt_cutoff} the resulting
lattice constant of CaF$_2$ is shown as a function of the cutoff
energy.

\section{splitting of energy levels of charge center due to symmetry
  considerations}
\label{sec:group}

In this appendix we give a few thoughts about symmetry, which are based on
group theoretical arguments. We assume that a charge center forms a
kind of hydrogen atom which is perturbed by the symmetry of the
underlying crystalline lattice. In the case of the $F$-center this is
obvious since the fluor vacancy forms a net positive charge in which
the excess electron is localized due to Coulomb attraction. The point
symmetry group of the $F$-center is tetrahedron symmetry, T$_d$. In the
case of the $H$-center this is less obvious. The excess fluorine
ion brings a negative charge which is compensated by a positive charge
which is localized among the four F-ions as shown in
fig. \ref{fig:fig6}. The point symmetry of the whole $H$-center is C$_{2v}$.

The optically allowed  transitions within the
hydrogen atom are given by well known rules which follow from symmetry
considerations in the rotation group, O(3)=R+C$_i$. The angular part of the
eigenfunctions in the O(3) group are the spherical harmonics,
D$_l^p$. Where $l$ is the angular index of the spherical harmonics and
$p$ the parity (+ or -) of the wave function. Under
the perturbation of the external field these (2l+1) times degenerate
eigenvalues split in a well defined way. These is given in correlation
tables which can be found in standard textbooks of group theory
\cite{hamermesh1990}. In table \ref{tab:corr_group} we give the 
part of the correlation table which is relevant for the symmetry of
the charge centers. From table   \ref{tab:corr_group} we learn the
following important things: (i) the splitting of the energy levels in
the excited states and second the numbers of independent parameters
which are required to expand the crystal field up to fourth order in
$l$. In the case of the $F$-center only the second excited state splits
into a twofold (E) and a threefold (T$_{2/1}$) degenerate energy
level. The third excited state splits into three levels, one highly
symmetric A$_{2/1}$ and two levels which are each threefold
degenerate T$_1$ and T$_2$. In the case of the $H$-center all
degeneracies up to fourth order are  canceled, already the
first excited state splits into three non degenerate energy levels  
A$_{2/1}$, B$_1$ and B$_2$. The second excited energy level splits into
five different levels and so on. By (ii) counting the number of A$_1$
representations we know the number of independent parameters which are
needed for an expansion of the crystal field up to fourth order in
$l$. For the $F$-center these are only 3 parameters whereas for the
$H$-center with its much lower symmetry 11 parameters are required.

\end{appendix}


\begin{thebibliography}{10}
\expandafter\ifx\csname bibnamefont\endcsname\relax
  \def\bibnamefont#1{#1}\fi
\expandafter\ifx\csname bibfnamefont\endcsname\relax
  \def\bibfnamefont#1{#1}\fi
\expandafter\ifx\csname url\endcsname\relax
  \def\url#1{\texttt{#1}}\fi
\expandafter\ifx\csname urlprefix\endcsname\relax\def\urlprefix{URL }\fi
\providecommand{\bibinfo}[2]{#2}
\providecommand{\eprint}[2][]{\url{#2}}

\bibitem{burnett2001}
\bibinfo{author}{\bibnamefont{{J.H.~Burnett, Z.H.~Levine, E.L.~Shirley}}},
  \bibinfo{journal}{Phys.~Rev.~B, {\bf 64} 241102R}  (\bibinfo{year}{2001}).

\bibitem{letz2003b}
\bibinfo{author}{\bibnamefont{{M.~Letz, L.~Parthier, A.~Gottwald,
  M.~Richter}}}, \bibinfo{journal}{{Phys.~Rev.~B {\bf 67}, 233101}}
  (\bibinfo{year}{2003}).

\bibitem{hayes1974}
\bibinfo{author}{\bibfnamefont{E.~W.} \bibnamefont{Hayes}},
  \emph{\bibinfo{title}{Crystals with the fluorite structure, pp 102}}
  (\bibinfo{publisher}{Clarendon Press, Oxford}, \bibinfo{year}{1974}).

\bibitem{fowler1968}
\bibinfo{author}{\bibfnamefont{W.B.~Fowler} \bibnamefont{(Editor)}},
  \emph{\bibinfo{title}{Physics of Color Centers}}
  (\bibinfo{publisher}{Academic Press, New York}, \bibinfo{year}{1968}).

\bibitem{tsujibayashi2000}
\bibinfo{author}{\bibnamefont{{T.~Tsujibayashi, M.~Watanabe, O.~Arimoto,
  M.~Itoh, S.~Nakanishi, H.~Itoh, S.~Asaka, M.~Kamada}}},
  \bibinfo{journal}{J.~of~Luminescence, {\bf 87-89}, 254}
  (\bibinfo{year}{2000}).

\bibitem{goerling2005}
\bibinfo{author}{\bibnamefont{{Ch.~G{\"o}rling, U.~Leinhos, K.~Mann}}},
  \bibinfo{journal}{Opt. Comm. {\bf 249}, 319}  (\bibinfo{year}{2005}).

\bibitem{tomiki69}
\bibinfo{author}{\bibnamefont{{T.~Tomiki, T.~Miyata}}},
  \bibinfo{journal}{J.~Phys.~Soc.~Jpn. {\bf 27}, 658}  (\bibinfo{year}{1969}).

\bibitem{barth90}
\bibinfo{author}{\bibnamefont{{J.~Barth, R.L.~Johnson, M.~Cardona, D.~Fuchs,
  A.M.~Bradshaw}}}, \bibinfo{journal}{Phys.~Rev.~B {\bf 41}, 3291}
  (\bibinfo{year}{1990}).

\bibitem{williams1976}
\bibinfo{author}{\bibnamefont{{R.~T.~Williams, M.~N.~Kabler, W.~Hayes,
  J.~P.~Stott}}}, \bibinfo{journal}{Phys.~Rev.~B, {\bf 14}, 725}
  (\bibinfo{year}{1976}).

\bibitem{lindner01}
\bibinfo{author}{\bibnamefont{{R.~Lindner, R.T.~Williams, M.~Reichling}}},
  \bibinfo{journal}{Phys.~Rev.~B {\bf 63}, 075110}  (\bibinfo{year}{(2001)}).

\bibitem{mizuguchi1999}
\bibinfo{author}{\bibnamefont{{M.~Mizuguchi, H.~Hosono, H.~Kawazoe}}},
  \bibinfo{journal}{J.~Opt.~Soc.~Am. B, {\bf 16}, 1153}
  (\bibinfo{year}{1999}).

\bibitem{tanimura2001}
\bibinfo{author}{\bibnamefont{{K.~Tanimura}}}, \bibinfo{journal}{Phys.~Rev.~B,
  {\bf 63}, 184303}  (\bibinfo{year}{2001}).

\bibitem{muehlig2002}
\bibinfo{author}{\bibnamefont{{Ch.~M\"uhlig, W.~Triebel, G.~T\"opfer,
  A.~Jordanov}}}, \bibinfo{journal}{Proc. Boulder damage symposium}
  (\bibinfo{year}{2002}).

\bibitem{cramer2005}
\bibinfo{author}{\bibnamefont{{L.P.~Cramer, T.D.~Cumby, J.A.~Leraas,
  S.C.~Langford, J.T.~Dickinson}}}, \bibinfo{journal}{J.~Appl.~Phys. {\bf 97},
  103533}  (\bibinfo{year}{2005}).

\bibitem{170K}
\bibinfo{author}{\bibfnamefont{E.~W.} \bibnamefont{Hayes}},
  \emph{\bibinfo{title}{Crystals with the fluorite structure, pp 240}}
  (\bibinfo{publisher}{Clarendon Press, Oxford}, \bibinfo{year}{1974}).

\bibitem{kuzokov1998}
\bibinfo{author}{\bibnamefont{{V.N.~Kuzokov, E.A.~Kotomin, W.~von~Niessen}}},
  \bibinfo{journal}{Phys.~Rev.~B {\bf 58}, 8454}  (\bibinfo{year}{1998}).

\bibitem{atobe1979}
\bibinfo{author}{\bibnamefont{{K.~Atobe}}}, \bibinfo{journal}{J.~Chem.~Phys.
  {\bf 71}, 2588}  (\bibinfo{year}{1979}).

\bibitem{vasp}
\bibinfo{author}{\bibnamefont{{G.~Kresse, J.~Furthm{\"u}ller}}},
  \bibinfo{journal}{Phys.~Rev.~B {\bf 54}, 11169}  (\bibinfo{year}{1996}).

\bibitem{kim2004}
\bibinfo{author}{\bibnamefont{{M.~Kim, Y..~Zhao, A.~J.~Freeman,
  W.~Mannstadt}}}, \bibinfo{journal}{Appl.~Phys.~Lett. {\bf 84}, 3579}
  (\bibinfo{year}{2004}).

\bibitem{schmalzl2003}
\bibinfo{author}{\bibnamefont{{K.~Schmalzl, D.~Strauch, H.~Schober}}},
\bibinfo{journal}{Phys.~Rev.~B, {\bf 68}, 144301} 
  (\bibinfo{year}{2003}).

\bibitem{khenata2005}
\bibinfo{author}{\bibnamefont{{R.~Khenata, B.~Daoudi, M.~Salnoun, H.~Batlache,
  M.~Rerat, A.H.~Reshak, B.~Bouhafs, H.~Abid, M.~Driz}}},
  \bibinfo{journal}{Eur. Phys. J. B {\bf 47}, 63}  (\bibinfo{year}{2005}).

\bibitem{bericht_spaeth}
\bibinfo{author}{\bibnamefont{{J.M.~Spaeth, S.~Schweizer}}},
  \bibinfo{journal}{internal report}  (\bibinfo{year}{2004}).

\bibitem{pick1972}
\bibinfo{author}{\bibfnamefont{H.}~\bibnamefont{Pick}},
  \emph{\bibinfo{title}{optical properties of solids, pp. 653}}
  (\bibinfo{publisher}{North-Holland publishing company},
  \bibinfo{year}{1972}).

\bibitem{shi2005}
\bibinfo{author}{\bibnamefont{{H.~Shi, R.I.~Eglitis, G.~Borstel}}},
  \bibinfo{journal}{Phys.~Rev.~B, {\bf 72}, 045109}  (\bibinfo{year}{2005}).

\bibitem{puchina1998}
\bibinfo{author}{\bibnamefont{{A.~V.~Puchina, V.~E.~Puchin, E.~A.~Kotomin,
  M.~Reichling}}}, \bibinfo{journal}{Sol.~Stat.~Com. {\bf 106}, 285}
  (\bibinfo{year}{1998}).

\bibitem{hamermesh1990}
\bibinfo{author}{\bibfnamefont{M.}~\bibnamefont{Hamermesh}},
  \emph{\bibinfo{title}{Group theory and its application to physical systems}}
  (\bibinfo{publisher}{Dover}, \bibinfo{year}{1990}).

\end{thebibliography}

\newpage

\section*{figure captions}

\begin{figure}[h]
\caption{band structure of CaF$_2$ from an LDA calculation. As
expected the band gap results in a too small value, whereas the main
features like e.g. the indirect transition are obtained in well
agreement with experiments.
}
\label{fig:fig1} 
\end{figure}
\begin{figure}[h]
\caption{band structure of Ca$_{32}$F$_{63}$ from an LDA calculation. 
A weakly dispersive defect level shows up in the band gap. the Fermi
level marked as $0$ of the energy scale is in the half filled defect
level indicating a free spin of the defect.  
}
\label{fig:fig2} 
\end{figure}
\begin{figure}[h]
\caption{Change in electronic density due to formation of an
  $F$-center. The $F$-center is well localized and accompanied by only a
  very small lattice distortion.
}
\label{fig:fig3} 
\end{figure}
\begin{figure}[h]
\caption{band structure of Ca$_{32}$F$_{63}$ from an LDA calculation. 
A weakly dispersive defect level shows up in the band gap. the Fermi
level marked as $0$ of the energy scale is in the half filled defect
level indicating a free spin of the defect.  
}
\label{fig:fig4} 
\end{figure}
\begin{figure}[h]
\caption{Change in electronic density due to formation of an
  H-center. The $H$-center is well localized and accompanied by a
  huge lattice deformation. The additional positive charge is located
  at the center of four F$^-$-ions forming an F$_4^{3-}$ complex. The
  symmetry of the $H$-center is C$_{2v}$. In appendix
  \protect{\ref{sec:group}} we estimate the level splitting due to
  symmetry arguments.
  }
\label{fig:fig5} 
\end{figure}

\begin{figure}[h]
\caption{The geometry of the 
 F$_4^{3-}$, $H$-center as it lies in the (-1,1,0) plane is shown. The
 angular as they result from the 
 calculation are plotted and given in $^o$ in fig. a). The distances between the
 four F--ions are given in \AA. The four F--ions form a perfect plane
 above and below which a Ca$^{2+}$--ion is located. The overall
 symmetry of the $H$-center obtained via our calculation is C$_{2v}$ as
 can be seen from fig. b).
}
\label{fig:fig6} 
\end{figure}

\begin{figure}[h]
\caption{Schematic plot of the energy levels of the $F$-$H$-pair before
  and after recombination. Resulting from the calculation there is a
  large part of the energy stored in a lattice deformation (Stokes
  shift). Therefore the measured 278nm fluorescence is smaller than
  the calculated energy difference of 8.17 eV.
}
\label{fig:fig7} 
\end{figure}

\begin{figure}
\caption{The total energy for CaF$_2$ as a function of plane wave
  cutoff. For cutoff energies of 500 eV the system starts to
  saturate. The upper (red) curve shows the result without structural
  relaxation and the lower (green) curve shows the result with
  structural relaxation of the elementary cell.
}
\label{fig:conv}
\end{figure}

\begin{figure}
\caption{The lattice constant of CaF$_2$ after structural relaxation
  is plotted as a function of plane wave cutoff.
}
\label{fig:latt_cutoff}
\end{figure}

\newpage

\section*{tables}

\begin{table}[h]
\begin{tabular}{|c|c|}
\cline{1-2}
type of  & $\Delta E_{M-aggl}$   \\
$M$-center & / eV  \\
\cline{1-2}
\cline{1-2}
$M$-100 & -0.85 \\
\cline{1-2}
$M$-110 & -0.14 \\
\cline{1-2}
$M$-111 (Ca) & -0.31 \\
\cline{1-2}
\end{tabular}
\caption{
Agglomeration energy for the formation of an $M$--center in
CaF$_2$. According to our result in the most favoured configuration
the two F--vacancies are oriented along the 100 direction. In this
direction the gain due to lattice relaxation is largest.
}
\label{tab:m-cent} 
\end{table}

\begin{table}[h]
\begin{tabular}{|c|c|c|c|c|c|c|c|}
\cline{1-8}
dopant & $\Delta E_{defect}$ & \multicolumn{2}{|c|}{$n_{eq}(T=300K)$}  &
\multicolumn{2}{|c|}{$n_{eq}(T=1500K)$}  & \multicolumn{2}{|c|}{$n_{eq}(T=1650K)$}  \\
& / eV  & / m$^{-3}$ & / ppm & / m$^{-3}$ & / ppm
& / m$^{-3}$ & / ppm \\
\cline{1-8}
\cline{1-8}
$F$-center & -1.35 & 2.49 10$^{4}$ & 9.74 10$^{-19}$ & 1.29 10$^{23}$ & 5.04 &
3.43 10$^{23}$ & 13.4 \\
\cline{1-8}
$H$-center & -6.82 & 1.75 10$^{-91}$ & 6.84
10$^{-114}$ & 1.20 10$^{4}$ &4.70 10$^{-19}$ & 1.72 10$^{6}$ & 6.72
10$^{-17}$ \\
\cline{1-8}
\end{tabular}
\caption{
For $F$- and $H$-center the equilibrium concentrations at different
temperatures are calculated. The results are given as well in absolute
concentrations [cm$^{-3}$] as well as in molar fractions [ppm].
}
\label{tab:tabfhconc} 
\end{table}

\begin{table}[h]
\begin{tabular}{|c|c|c|}
\cline{1-3}
&$F$-center & $H$-center\\
O(3)=R x C$_i$ & T$_d$ & C$_{2v}$\\
\cline{1-3}
\cline{1-3}
D$_0^{\pm}$ & A$_{1/2}$ & A$_{1/2}$ \\
\cline{1-3}
D$_1^{\pm}$ & T$_{1/2}$ & A$_{2/1}$+B$_1$+B$_2$ \\
\cline{1-3}
D$_2^{\pm}$ & E+T$_{2/1}$ & A$_{1/2}$+2B$_1$+2B$_2$ \\
\cline{1-3}
D$_3^{\pm}$ & A$_{2/1}$+T$_1$+T$_2$ & A$_{1/2}$+2A$_{2/1}$+2B$_1$+2B$_2$ \\
\cline{1-3}
D$_4^{\pm}$ & A$_{1/2}$+E+T$_1$+T$_2$ &
3A$_{1/2}$+2A$_{2/1}$+2B$_1$+2B$_2$\\
\cline{1-3}
....& .... & ....\\ 
\cline{1-3}
\end{tabular}
\caption{
For the first four excitations of a charge center we give the relevant
part of the correlation table for a splitting of the energy levels
with respect to the symmetry of the $F$- and the $H$- center.
}
\label{tab:corr_group}
\end{table}

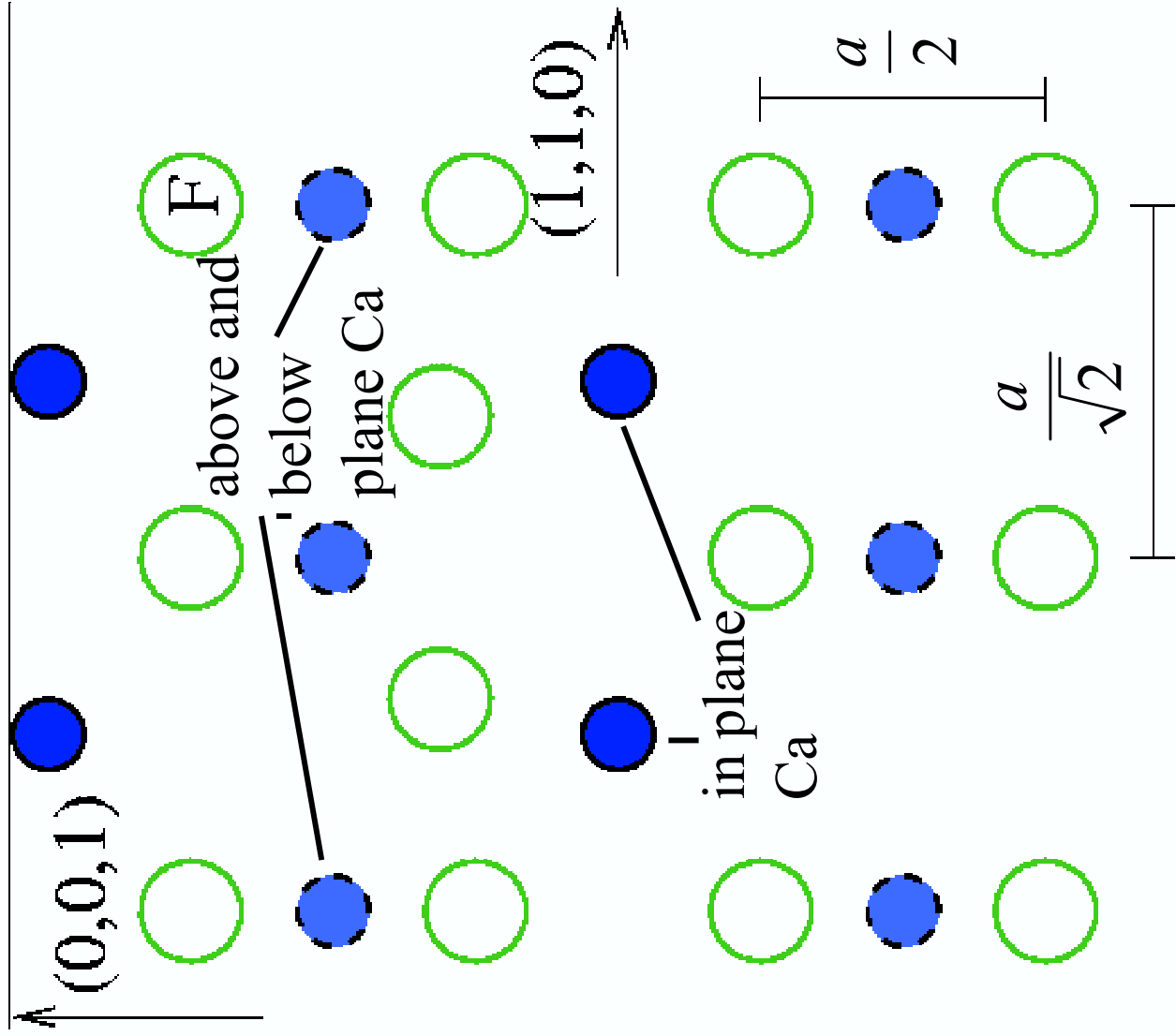
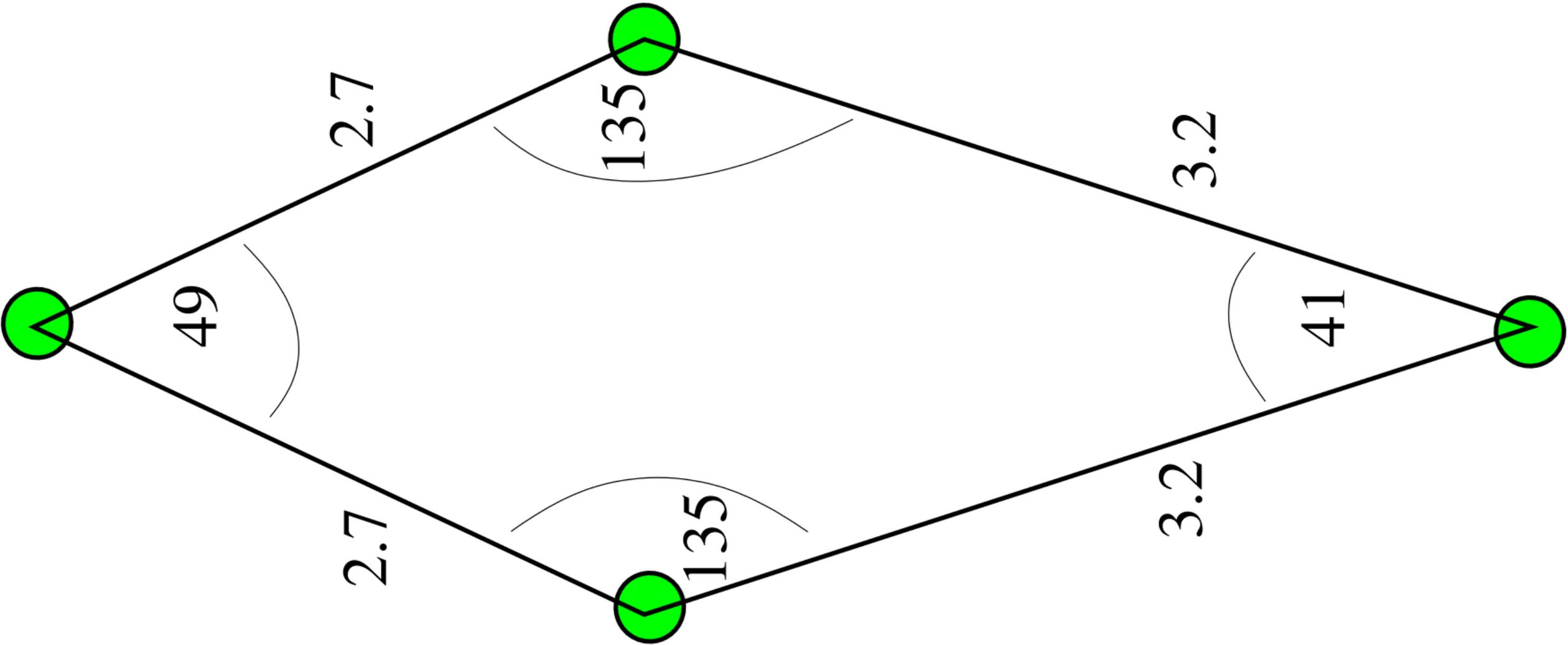
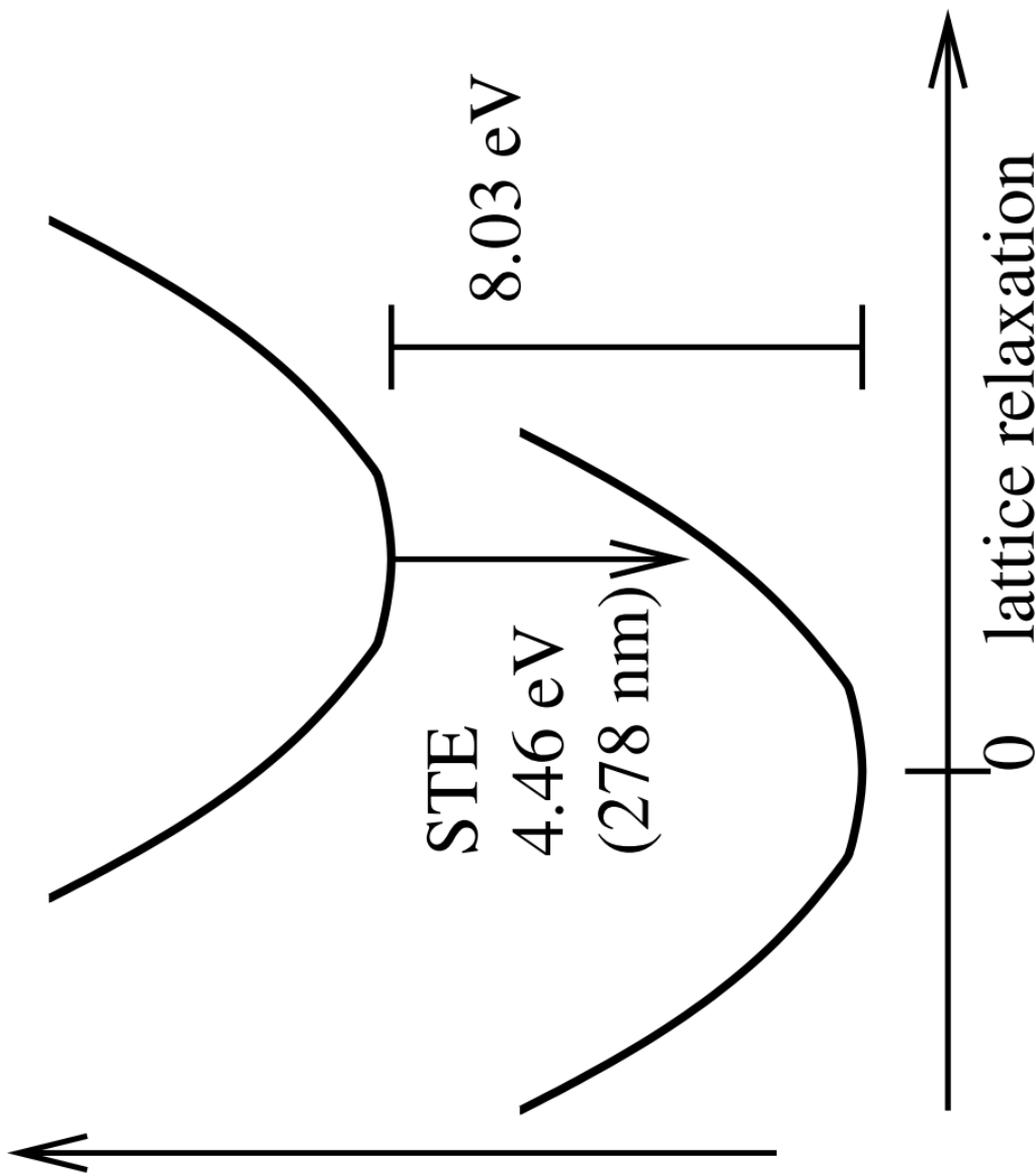
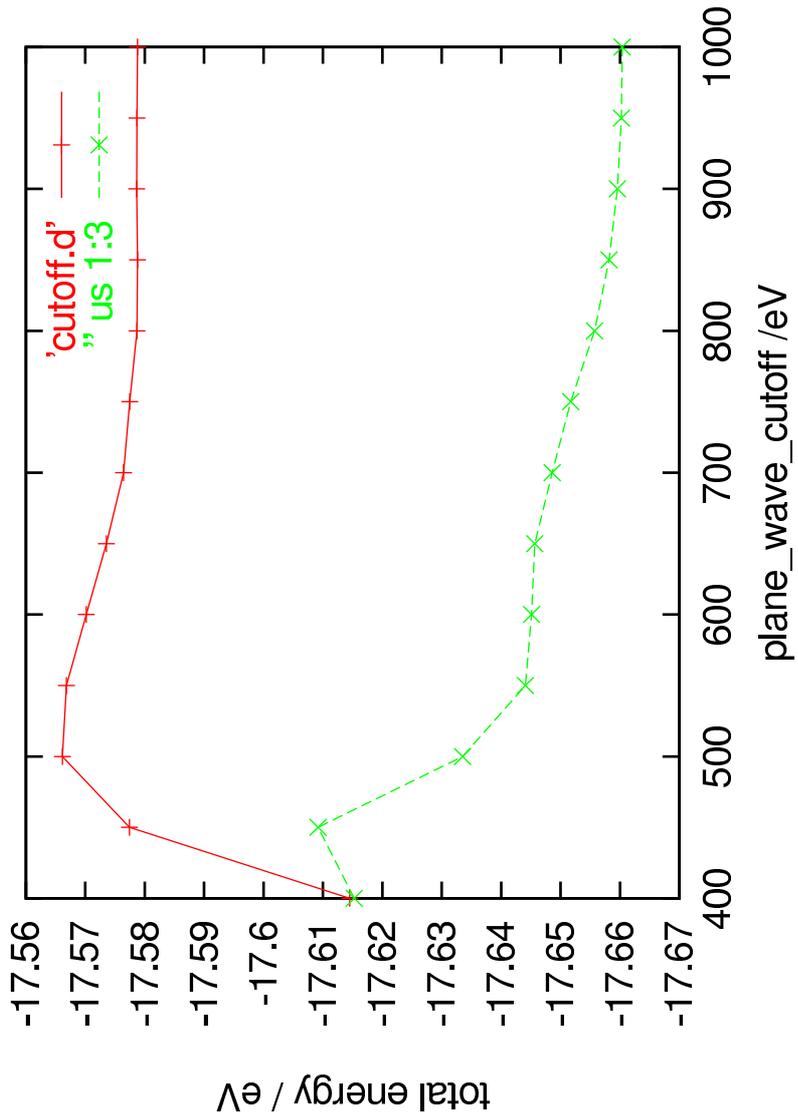
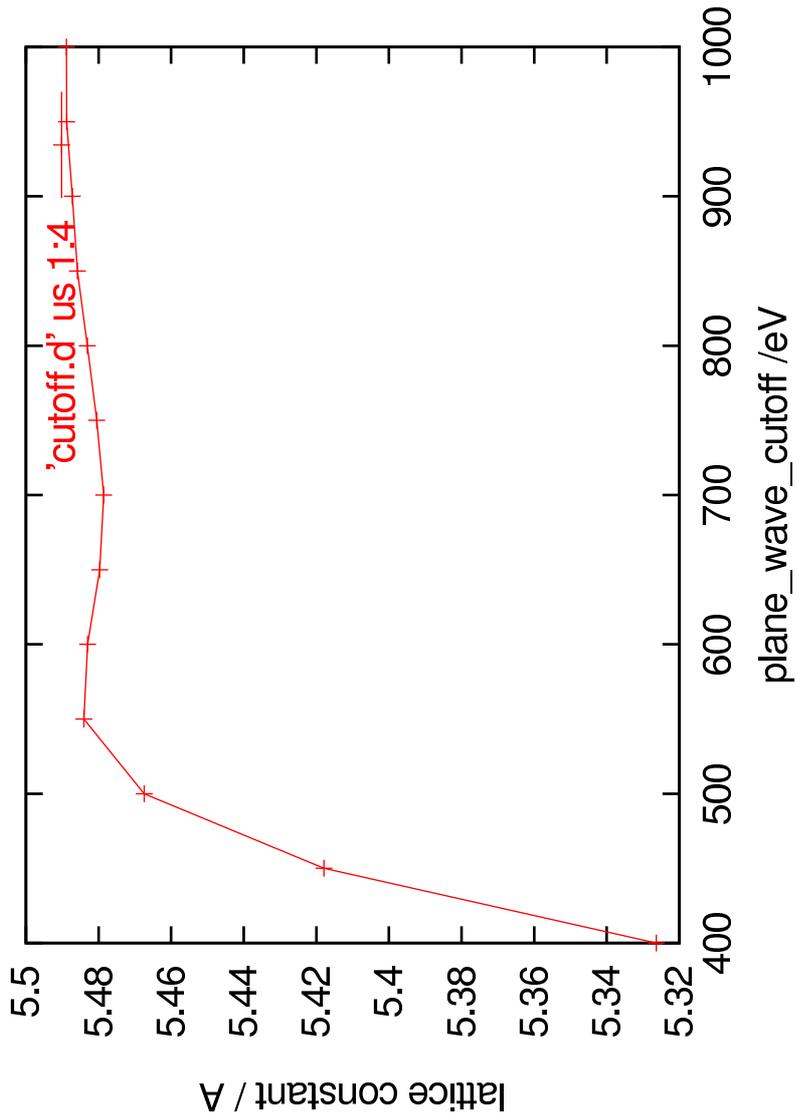
\end{document}